# AMERIGO VESPUCCI AND THE DISCOVERY OF THE SOUTHERN SKY

**Davide Neri**
*Liceo A.B.Sabin, Via Giacomo Matteotti 7, 40129 Bologna, Italy.*
E-mail: davideneri100@gmail.com

**Abstract**: During the voyages that led him to discover the new continent bearing his name, Amerigo Vespucci made interesting astronomical observations of the Southern sky. In the past, his data have been interpreted with criteria that do not follow Vespucci's indications, resulting in identifications that are not credible or even leading to the assertion that the data themselves are incomprehensible. However, it is possible to construct a coherent picture of all the information, arriving at an identification that is in some cases very probable, in other cases almost certain, of the stars described by Vespucci. Analysis of documents shows that he made good-quality measurements, but his incomplete knowledge of ancient texts prevented him from distinguishing the new stars from the already known ones, giving rise to a period of confusion in sixteenth century celestial cartography.

## 1 INTRODUCTION

Amerigo Vespucci (1454–1512) was the first European navigator to recognize, in some lands discovered between the late fifteenth and early sixteenth centuries, a new continent unknown to the ancient Greek and Roman geographers who provided the basis for the geographical knowledge of Europeans until the Age of Exploration. For this reason, Martin Waldseemuller, in his 1507 world map titled *Universalis Cosmographia*, attributed the name America to the land that we now call South America. The name was later extended to include lands further North, which had initially been considered part of Asia. The initial mistake arose from the fact that in those years there were two main hypotheses regarding the circumference of the Earth: it was about 40,000 km for Eratosthenes, about 30,000 km for Ptolemy. According to the Ptolemaic hypothesis it was possible to make a voyage from Spain to Japan, which was supposed to be where Mexico actually is, and when Cristoforo Colombo, with three Spanish ships, reached the Bahamas Islands, Cuba, and Hispaniola for the first time in 1492, everyone assumed he had reached Asia.

Vespucci was born in Florence, one of the richest and most culturally advanced cities in Europe at the end of the fifteenth century, and lived there at the same time as figures such as Leonardo da Vinci, Michelangelo Buonarroti, and Niccolò Machiavelli. He had a humanistic and technical education, which included knowledge of Latin (the language used by the ruling classes throughout Europe) and technical subjects, including geography, astronomy, and economics. These skills enabled him to pursue independent activities and be a useful collaborator for bankers, great merchants, and even governments.

Until 1503, he worked for Lorenzo di Pierfrancesco de Medici (1463–1503), a member of the important Florentine family that ruled the city from 1434 to 1494 and again from 1512 onward. Lorenzo sent him to Seville in 1492 to manage his financial and commercial activities related to the early voyages of exploration, but Vespucci also appears to have conducted business on his own. In the following years, he made at least two voyages to America on behalf of the kings of Spain and Portugal, likely acting as a pilot and cartographer.

Vespucci's role in the history of geographical knowledge is generally known. It is less well known that his true ambition, when he embarked on his voyages, was to be the first cartographer capable of charting the stars of the Southern sky that could not be seen from Europe due to the curvature of the Earth. Since ancient times, the existence of these stars has been known to astronomers and geographers, and the awareness of their existence was transmitted continuously in the Middle Ages through the writings of Aratus, Cicero, Pliny, Macrobius, Martianus Capella, and other authors. However, only with the Age of Exploration, which took European navigators South of the Equator, did their observation become possible. Three documents by Vespucci are known to contain astronomical information. They are letters addressed to Lorenzo di Pierfrancesco de Medici:

[A] A letter in Italian sent from Seville in July 1500 (Vespucci, 1984a: 51);
[B] A letter in Italian sent from Lisbon in 1502 (Vespucci, 1984b: 75);
[C] The letter *Mundus Novus* (*The New World*), printed in Latin in 1504, but probably written in the first half of 1503 before the recipient's death (Vespucci, 1504).

Although there are various texts attributed to Vespucci whose authenticity is questionable, it is very likely that documents [A], [B], and [C] were written by him: Text [C] was published and gave Vespucci fame as the discoverer of the new continent during his lifetime. It is difficult to believe it was the product of another author. The two preceding documents [A] and [B] are generally consistent with text [C] and provide a private anticipation of it. However, there are other texts about which serious doubts arise: in particular, two Italian translations of the *Mundus Novus* (Montalboddo, 1507) and (Ramusio, 1550), poorly reproduce both the paragraphs dedicated to astronomy and the related figures. These versions, which were then translated into other languages, may have prevented a correct interpretation of the original text for a long time (see Section 13 for a partial comparison).

Comparing the texts [A], [B], [C] with each other and with data derived from a map of the Southern sky dated to the year 1500 it is possible to provide many interesting insights into the quality of Vespucci's observations.

## 2 TEXT [A]: THE JULY 1500 LETTER FROM SEVILLE AND THE 'SPANISH EXPEDITION'

The document Vespucci (1984a: 51) contains information about a voyage undertaken with a Spanish expedition in 1499–1500. In the letter Vespucci states that he reached latitude 6° S and thus observed the South Celestial Pole (SCP) 6° above the horizon, immediately realizing the difficulty of locating it due to the lack of bright stars with a distance of less than 10° from it. Unable to imagine that his name would be associated with a new continent a few years later, he demonstrates here that his initial ambition was to be the first cartographer of the Southern sky. In the letter, he writes:

> I, wishing to be the discoverer of the south celestial pole, lost sleep many times at night, trying to note which star had the least movement and which was closest to the pole. But despite all the bad nights I had and with all the instruments I used (the quadrant and the astrolabe and other things to mark the stars), I could not identify a star that was less than 10 degrees away from the pole; so I was not satisfied in myself to name any of them as the south pole, because of the great circle that these stars made around it. (Vespucci, 1984a: 59).

Vespucci says nothing more about the instruments (astrolabe and quadrant) that he used.

In the sequel Vespucci speaks of four stars arranged in an "almond shape" (Vespucci, 1984a: 60), which he connects with some verses by Dante Alighieri, a Florentine poet who lived two centuries earlier. Dante, who in a popular treatises (the *Convivio*) emphasized the impossibility of observing the stars near the SCP due to the curvature of the Earth, in his poem entitled *Commedia* describes a fantastic journey through Hell, Purgatory, and Paradise. After traversing Hell inside the Earth with the guidance of the Latin poet Virgil, Dante returns to the surface at the antipodes of Jerusalem, and from there he observes four stars in the Southern Hemisphere. Vespucci, who like Dante was a Florentine and well-versed in the *Commedia*, hypothesizes that the four almond-shaped stars, which likely correspond to the Southern Cross (Crux), could be those imagined by Dante.

## 3 TEXT [B]: THE 1502 LETTER FROM LISBON ABOUT THE 'PORTUGUESE EXPEDITION'

In this letter, Vespucci (1984b: 75) provides information on a second voyage he undertook in 1501–1502 on behalf of the King of Portugal. The relevant cartographic and astronomical information is twofold: the expedition reached as far as 50° S, and the Southern sky revealed itself to be populated by "… very clear and beautiful stars." (Vespucci, 1984b: 78). Nothing else is said, because the results of the observations were recorded in a document that was delivered to the King, which Vespucci was hoping to get back for publication.

## 4 TEXT [C]: *MUNDUS NOVUS*, THE NEW CONTINENT AND THE ASTRONOMICAL OBSERVATIONS

In both [A] and [B] Vespucci considers the lands he explored on his two voyages to be the easternmost reaches of Asia. His belief that he had made a great geographical discovery was announced in *Mundus Novus* (Vespucci, 1504). The text, a letter in Latin of just a few pages, was printed in 1504 and was a great success. It begins by declaring that the voyage of 1501–1502 (already described in private letter [B]) led to the discovery of a new world, "... because our ancestors had no knowledge of them, and it will be a matter wholly new to all those who hear about them." (Vespucci, 1916: 1). Then the statement, already written in text [B], that the expedition reached the latitude of 50° S follows. Therefore, Vespucci was able to observe the Southern sky in optimal conditions. The text contains various astronomical details, which I have divided into six points [1]–[6]. Here, they are presented in the original Latin version (Vespucci, 1504) and then as English translations (Vespucci, 1916), where I have inserted some points of clarification to facilitate the interpretation of the text.

## 5 POINT [1]: THE SOUTHERN SKY IS POPULATED BY NUMEROUS VERY BRIGHT STARS

Vespucci (1504): Celum speciosissimis signis et figuris ornatum est, in quo annotavi stellas circiter viginti tante claritatis quante aliquando vidimus Venerem et Iovem. Harum et motus et circuitionem consideravi …

Vespucci (1916): The sky is adorned with most beautiful constellations and forms among which I noted about twenty stars as bright as we ever saw Venus or Jupiter. I have considered the movements and orbits of these ...

Here Vespucci claims to have measured the distances of the brightest stars from the SCP.

## 6 POINT [2]: THREE NEBULAE, CALLED 'CANOPI' (SINGULAR: CANOPUS), CAN BE OBSERVED

Vespucci (1504): Vidi in eo celo tres canopos, duos quidem claros, tertium obscurum.

Vespucci (1916): I saw in that sky three canopi, two indeed bright, the third dim.

The bright "canopi" are easily identifiable as the Large Magellanic Cloud (LMC) and the Small Magellanic Cloud (SMC); the dark one with the Coalsack Nebula (CN), a dark nebula that stands out against the bright background of the Milky Way. The identification of the dark canopus with CN will become evident on the basis of the description of the stars that are close to it.

## 7 POINT [3]: THERE ARE NO BRIGHT STARS NEAR THE SOUTH CELESTIAL POLE

Vespucci (1504): Polus antarticus non est cum Ursa Maiore et Minore, ut hic noster videtur articus, nec iuxta cum conspicitur aliqua clara stella …

Vespucci (1916): The Antarctic [celestial] pole is not figured with a Great and a Little Bear as this Arctic pole of ours is seen to be, nor is any bright star to be seen near it ...

Vespucci reiterates here what he stated in letter [A]: near the SCP, there are no bright stars that allow it to be easily identified.

## 8 INTERPRETATION: PART 1

Points [1], [2] and [3] show that Vespucci excluded stars of modest brightness from his interest and clearly indicates his criterion for representing the Southern sky, which focuses on celestial objects most easily identifiable by a navigator, even in less than optimal meteorological conditions. This seems to undermine any attempt to interpret Vespucci's text by considering stars of magnitude greater than 4 or even 5, like the one found in (Houzeau, 1885–1886). As will become evident in Section 13, Houzeau was misled because he did not use the original version of the *Mundus Novus*, but the one found in Ramusio (1550), which introduces significant distortions in the astronomical part of the text.

British amateur astronomer Edward Ball Knobel (1917: 415), for his part, also seems to have misunderstood the meaning of Vespucci's descriptions, coming to the conclusion that the stars described by him "... cannot be identified." This is true if one looks for faint stars close to the SCP, but it is not true if one looks for bright stars more than 10° from it, as indicated by Vespucci himself. Following this last criterion, in my analysis of the points [4], [5] and [6] I will consider stars with $m_V$<3.5, because the stars it uses as its main references (and which can be uniquely identified by their distance from the pole) are brighter than this magnitude. The objects to be considered on the basis of these indications are shown in Figure 1, which refers to the Southern sky up to a distance of about 35° from the SCP.

The black cross indicates the SCP in the year 1500, and the black circles (●) are stars with $m_V$<3.5. The LMC, SMC, and CN are the three canopi described by Vespucci. In addition, some faint stars belonging to the constellations of Chamaeleon and Apus are reported with a white color and black border (○). They have been included only because their distance from the SCP is about 12.5° and will be useful in the analysis of point [5]. Another star represented with a white color and black border is η Carinae, which is a variable star with a current magnitude $m_V$ = 6.2, but which in 1845 had reached a magnitude $m_V$ –1.0. The irregularity of its behavior does not allow us to know how bright it was in 1501–1502 (Smith and Frew, 2011).

Before analyzing these points, however, it is appropriate to consider some fundamental aspects of the observation of the SCP by an observer located in the Southern Hemisphere. The stars surrounding the SCP rotate around it in a clockwise direction, unlike the stars in the Northern Hemisphere, which rotate counterclockwise around the NCP. The difference arises from the fact that, in both cases, the stars rise in the East and set in the West, but for an observer looking South, East is on the left, while for an observer looking North, East is on the right. Taking this into account, we can make sense of Vespucci's statements that one group of stars "follows" another, referring to their clockwise rotation around the SCP.

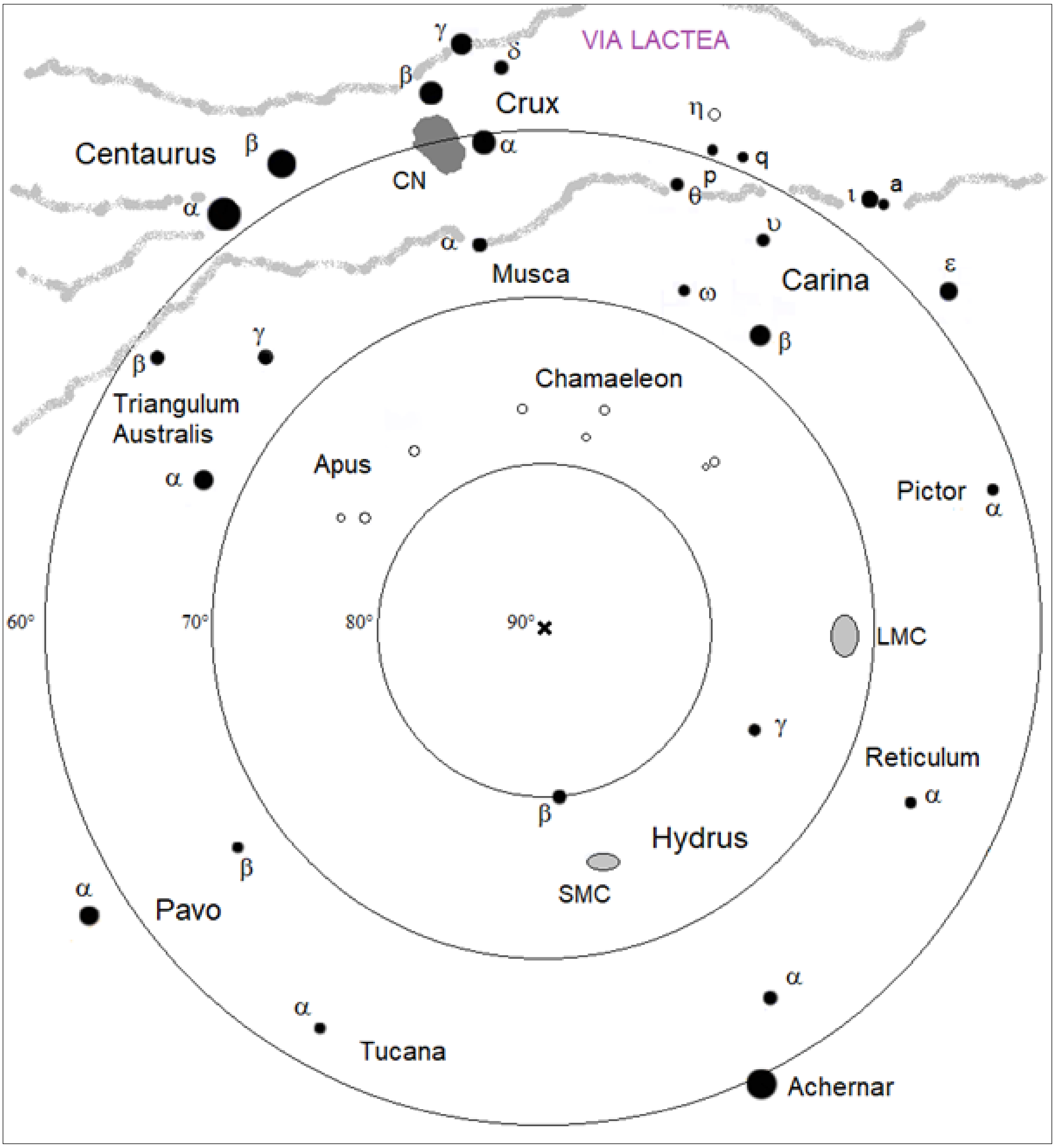

Figure 1: The South polar region within 35° of the Pole in 1500 (plot: the author).

In Figure 1 it is evident that, as Vespucci states, there are no bright stars near the SCP, but many are found at greater distances, especially in the area occupied by the current constellations of Crux (the Southern Cross), Triangulum Australis (the Southern Triangle), Centaurus and Carina.

The information in the following points [4]–[5]–[6] have a common feature: each of them refers to a different canopus and to stars whose distance from the SCP is indicated. The three points should be understood as independent of each other and must be referring to different objects, therefore in my opinion they should be interpreted separately. Points [4] and [6] are also accompanied in *Mundus Novus* by two figures, in which the canopi are represented by shapes obtained with typographical signs. The reading of these figures must be carried out taking into account that their accuracy could have been sacrificed for printing needs, worsening the understanding of the information contained in the text.

## 9 POINT [4]: THREE STARS WITH A BRIGHT CANOPUS – DISTANT 9.5° FROM THE SOUTH CELESTIAL POLE

Vespucci (1504): … ex hiis que circum eum cum breviore circuitu feruntur tres sunt habentes trigoni orthogoni schema, quarum dim-idia peripherie diametrus gradus habet novem semis. Cum hiis orientibus a leva conspicitur unus canopus albus eximie magnitudinis, que cum ad medium celum perveniunt hanc habent figuram.

Vespucci (1916): … of those [stars] which move around it [the SCP] with the shortest circuit there are three that have the form of an orthogonal [right-angled] triangle, the half circumference [of the motion of these stars], the diameter, has nine and a half degrees. Rising with these to the left is seen a white canopus of extraordinary size which when they reach mid-heaven have the form depicted in Figure 2.

Here Vespucci is talking about three stars forming a right-angled triangle, which is distant approximately 9.5° from the SCP. They are located near a bright canopus. Since there are no three stars that are all 10° from the SCP and form a right-angled triangle, the distance indicated must be that of the closest star, which is taken as a reference for the distance of the entire triangle. On the map, there is only one star with this characteristic: β Hydri ($m_V$ = 2.80). The “right-angled triangle” must therefore have β Hydri at one of its vertices. Observing the map, we note that β Hydri can form three different ‘quasi’ right-angled triangles, which are shown in Figure 3. The position of β Hydri leads to the identification of the bright canopus referred to by Vespucci with the SMC, because the LMC is too distant and lies beyond γ Hydri, a star that appears to be described in the following point [5] in relation to this second bright canopus.

- The first triangle is formed by β Hydri, α Hydri ($m_V$ = 2.86), and α Tucanae ($m_V$ = 2.90). The spherical angle at the vertex in β Hydri is 89°. Relative to this triangle, the SMC lies within the right angle, although its position does not coincide with that shown in Vespucci’s first figure (Figure 2).
- The second is formed by β Hydri, β Pavonis ($m_V$ = 3.42), and α Trianguli Australis ($m_V$ = 1.91). However, two characteristics rule it out: first, the spherical angle at the vertex in β Pavonis is approximately 83°; second, the position of the SMC with respect to this triangle is external and is in stark contrast to the Figure 2.
- The third triangle is formed by α, β and γ Hydri ($m_V$ = 3.26), but this time the spherical angle with vertex in γ Hydri is only 79°. However, this triangle corresponds better than the others to Figure 2, with the SMC placed near β Hydri. This correspondence has led Omodeo (2020) to consider it the triangle indicated by Vespucci, but the fact that it involves γ Hydri suggests that we exclude it. In fact, this star is almost certainly the one Vespucci refers to in following point [5], which must be interpreted independently of point [4], because they clearly refer to different groups of celestial objects.

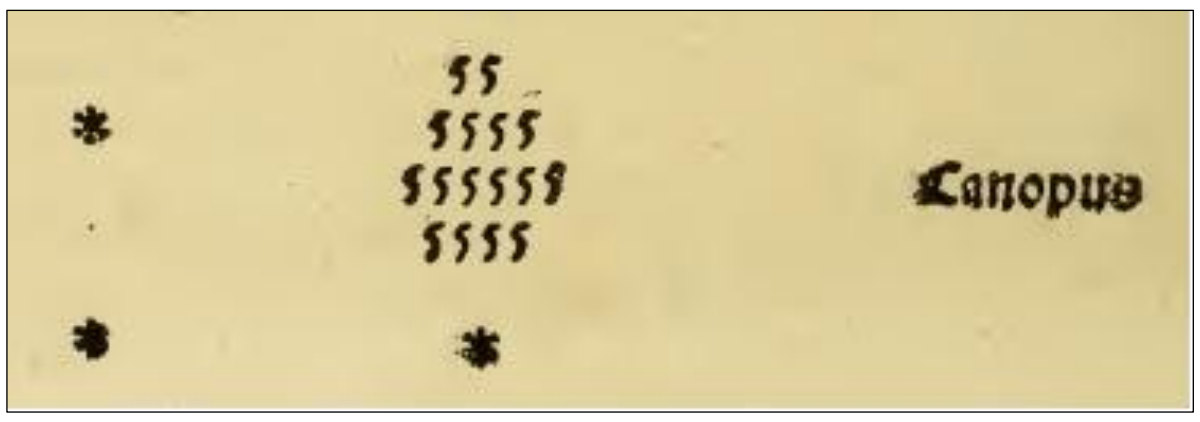

Figure 2: The first figure in *Mundus Novus* (1504) (https://archive.org/details/mundusnouus00vesp, courtesy: John Carter Brown Library).

The joint interpretation of points [4] and [5] will lead us to consider the triangle formed by β Hydri, α Hydri and α Tucanae as the one described by Vespucci.

## 10 POINT [5]: TWO STARS WITH A BRIGHT CANOPUS – DISTANT 12.5° FROM THE SOUTH CELESTIAL POLE

Vespucci (1504): Post has veniunt alie due, quarum dimidia peripherie diametrus gradus habet duodecim semis et cum eis conspicitur alius canopus albus.

Vespucci (1916): After these come two others [stars], the half circumference of which, the diameter [of their motion around the SCP], has twelve and a half degrees; and with them is seen another white canopus.

There is mention of two stars, one of which is 12.5° from the SCP. They are associated with the other bright canopus, which must the LMC,

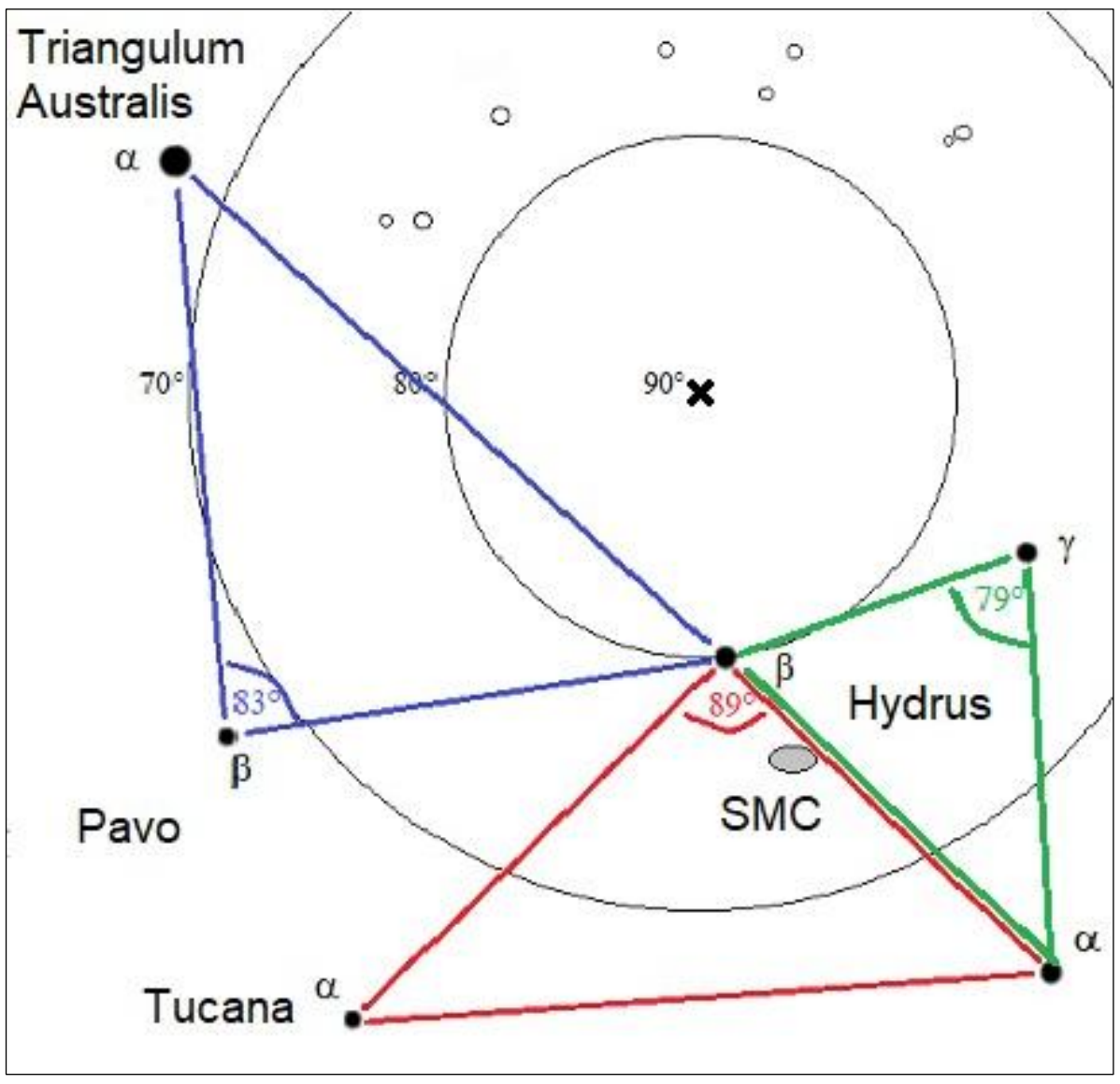

Figure 3: The three right-angle triangles with a vertex at β Hydri (plot: the author).

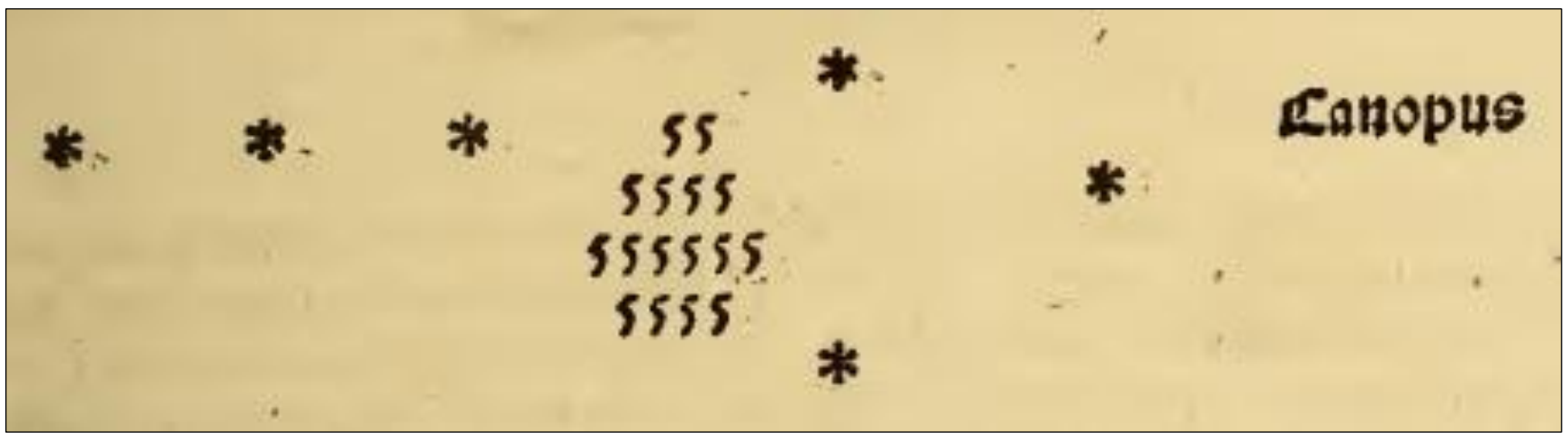

Figure 4: The second figure depicted in Mundus Novus (1504) (https://archive.org/details/mundusnouus00vesp, courtesy: John Carter Brown Library).

since the SMC is included in group [4]. The star that is approximately 12.5° from the SCP is most likely γ Hydri. As I highlighted in Figure 1, there are other faint stars at a similar distance from the SCP, five in the constellation of Chamaeleon, with $4.0 < m_v < 4.4$, and three in Apus, with $3.8 < m_v < 3.9$. However, these eight fainter stars are distant from the bright canopi, while the star cited by Vespucci must be close to the LMC. The lack of a figure leaves uncertainty as to which is the second star indicated by Vespucci. It is more probably α Reticuli ($m_v = 3.33$), less probably α Pictoris ($m_v = 3.27$). This group of stars follows the triangle of point [4] in its clockwise rotation around the SCP, as Vespucci said, and the structure of the sentence suggests that the LMC follows the two stars in their rotation.

## 11 POINT [6]: SIX STARS WITH A DARK CANOPUS – DISTANT 32° FROM THE SOUTH CELESTIAL POLE

Vespucci (1504): Hiis succedunt alie sex stelle formosissime et clarissime inter omnes alia octave sphere, que in firmamenti superficie dimidiam habent peripherie diametrum graduum triginta duorum. Cum hiis pervolat unus canopus niger immense magnitudinis. Conspiciuntur in Via Lactea et huiusmodi figuram habent quando sunt in meridionali linea:

Vespucci (1916): There follow, upon these, six other most beautiful stars and brightest among all the others of the eighth sphere, which in the upper firmament have a half circumference, a diameter, of thirty-two degrees. With them revolves a black canopus of huge size.

They are seen in the Milky Way and have a form like this when observed on the meridian line. The interpretation of this point seems to be easy. There is a general agreement among scholars that the six stars shown in Figure 4 with the dark canopus (CN) are α and β Centauri and α, β, γ and δ Crucis. See, for example, Omodeo (2020) and Warner (1979). The distance of the group from the SCP is approximately 32°. The dark nebula is shown in the figure in the correct position, as can be seen by comparing Figure 4 and Figure 1.

## 12 INTERPRETATION - PART 2

Figure 5 following, highlights the three groups of celestial objects described in points [4]-[5]-[6] of the *Mundus Novus* identified with the interpretation presented above. It can be noted that group [5] follows, in the clockwise rotation around the SCP, group [4], and group [6] follows [5], as stated by Vespucci. Considering the three values of distances indicated by Vespucci, it seems that he made good quality distance measurements and that he identified the position of the SCP with good precision.

At the end of the astronomical part of *Mundus Novus* Vespucci confirms what he had already written in letter [B]: further information is contained in a short book that he delivered to the King of Portugal and hoped he would have it back soon. The fact that Vespucci never published other texts with more detailed astronomical information suggests that this text was never returned to him, and nor is it mentioned in other documents.

## 13 CONFUSION IN SIXTEENTH-CENTURY CARTOGRAPHY OF THE SOUTHERN SKY

Following Vespucci, sixteenth-century cartographers did not recognize the six stars of his second figure (Figure 4) as already known stars. However, they had been listed by Ptolemy in the *Almagest* catalogue with numbers from 31 to 36 in the constellation of Centaurus, where α and β correspond to the front legs, and the other four stars to the hind legs of the mythological creature (see Toomer, 1998). In ancient times, it was possible to see them very low on the horizon at the latitude of Rhodes (36° N). Subsequently, due to the precession of the equinoxes, they became less and less visible because their declination dropped to about 60° S, as reported by Vespucci, who saw them at about 30° from the SCP.

As already mentioned, the inaccuracies of the Italian translations of Vespucci's Latin text contributed to the distortion of the available information. The differences between the translations and the original are particularly evident in the sections of *Mundus Novus* dedicated to astronomy, and seem to indicate that the translators lacked sufficient expertise to understand the text.

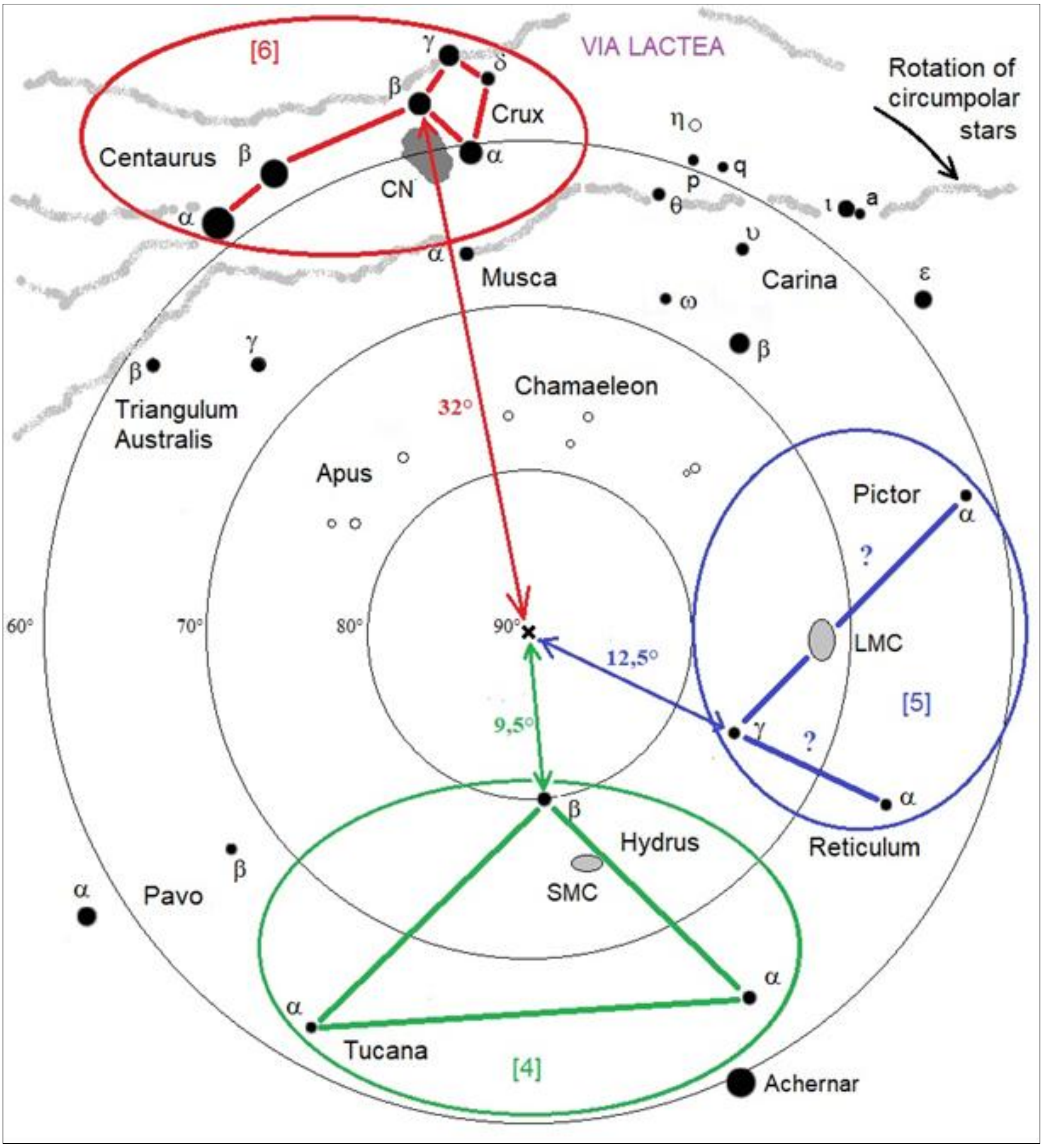

Figure 5: General interpretation of points [4], [5] and [6] (diagram: the author).

I will limit myself here to giving an example that concerns points [3] and [4], comparing only Vespucci (1504) and Ramusio (1550), because the diffusion and modifications of these documents require a more extensive study. Here is what *Mundus Novus* (Vespucci, 1504) says:

The Antarctic [celestial] Pole is not figured with a Great and a Little Bear as this Arctic Pole of ours is seen to be, nor is any bright star to be seen near it, and of those [stars] which move around it [the Pole] with the shortest circuit there are three which have the form of an orthogonal [right-angled] triangle, the half circumference [of the motion of these stars], the diameter, has nine and a half degrees. Rising with these to the left is seen a white canopus of extraordinary size which when they reach mid-heaven have this form [Figure 6]:

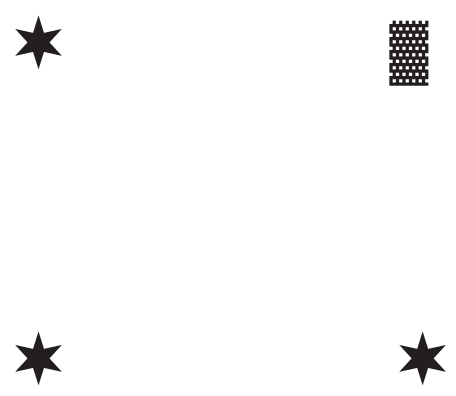

Figure 6

Comment: The ▓ is a canopus.

Meanwhile, the Italian translation in Ramusio (1550) says:

> The Antarctic [celestial] Pole does not have a Big Bear or a Little Bear, as can be seen at our Arctic Pole, nor do any shining stars touch it, but those that surround it are four, which have the shape of a quadrangle [see Figure 7].

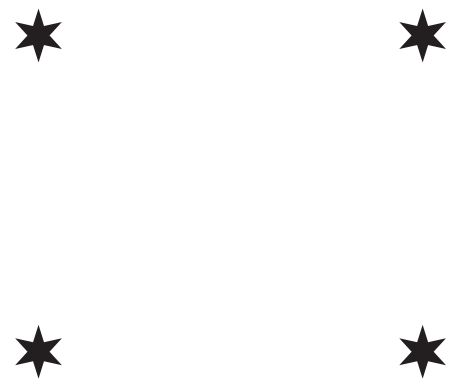

Figure 7

> And while these are rising, a resplendent canopus of notable size can be seen on the left side, which, having risen to the middle of the sky, represents the undersigned figure [Figure 8].

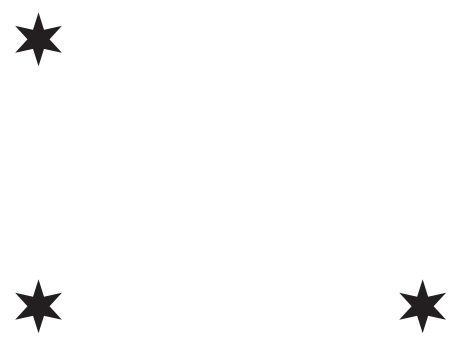

Figure 8

Comment: nothing is said about the star which is 9.5° from the SCP. There are two figures instead of one. Figure 7 seems to be Figure 6, but with a star instead of the canopus; Figure 8 still seems to be Figure 6 but without the canopus. Anyway, the Latin edition of 1504 never says that there are four stars that 'surround' the SCP. Finally, the 'quadrangle' of the text is represented in Figure 7 as a rectangle. It should be noted that the first English translation of the *Mundus Novus*, published in *Decades of Newe Worlde* (Eden, 1555), comes from (Ramusio, 1550). In his paper Houzeau (1885–1886) refers to Ramusio, and identifies the four stars of Figure 7 in ζ ($m_v$ = 5.42), η ($m_v$ = 6.19), κ ($m_v$ = 5.55) and σ ($m_v$ = 5.42) Octantis and the three stars in Figure 8 as γ ($m_v$ = 4.12), δ ($m_v$ = 5.46) and ζ ($m_v$ = 5.11) Chamaeleontis, without any reference to the Magellan Clouds. It is inevitable to point out that it is easier to find the four stars indicated by Houzeau when the position of the SCP is already known, than to find the SCP by first identifying four stars with $m_v$>5.

After Vespucci, the situation became even more confusing in 1516, when Andrea Corsali returned from a trip to India and published a global map of the Southern polar region marred by numerous inaccuracies (Corsali, 1516). The analysis of the Corsali map and its influence on the history of cartography of the Southern sky will be the subject of a paper currently in preparation.

The lack of alignment between old and new knowledge, which lasted until the end of the century, meant that 'new' stars were added to ancient maps in arbitrary positions, making the early cartography of the Southern sky very imprecise. The situation was clarified only with the data collected by the Dutch Keyzer–Houtmann expedition of 1595–1597 and the new celestial globe published by Petrus Plancius in 1598, where he introduced twelve new constellations and gave a coherent description of the Southern sky. On this development of cartographic knowledge see, for example, de Grijs (2023).

Meanwhile, for sixteenth-century navigators the Southern Cross had become the primary instrument for orienting themselves in the Southern sky. To locate the SCP, Pedro de Medina (1545), correctly suggested drawing a straight line starting from γ Crucis, passing through α Crucis, and continuing for another 30° (see Figure 1 and Figure 5).

To appreciate the inaccuracy of maps prior to 1598, we can take for example a small Southern Hemisphere inserted by Plancius himself in his world map *Orbis Terrae Compendiosa Descriptio* of 1596 (Figure 9), only one year before F. de Houtman brought to Europe much more precise data collected with P. Keyser on the positions of the Southern stars. Since the map is centered on the Ecliptic Pole, the SCP is not in the center but further to the left, and is identified by the white circle. It is easy to see that the Southern Cross is positioned so that α Crucis is 30° from the SCP, but it is far from Centaurus, and its size is overestimated.

## 14 CONCLUDING REMARKS

Throughout the sixteenth century, celestial cartography of the Southern sky in Europe underwent a pre-scientific phase, characterized by the following elements:

- Observations were made by people with basic astronomical knowledge, lacking the professional skills needed to reconcile new observations with existing knowledge.

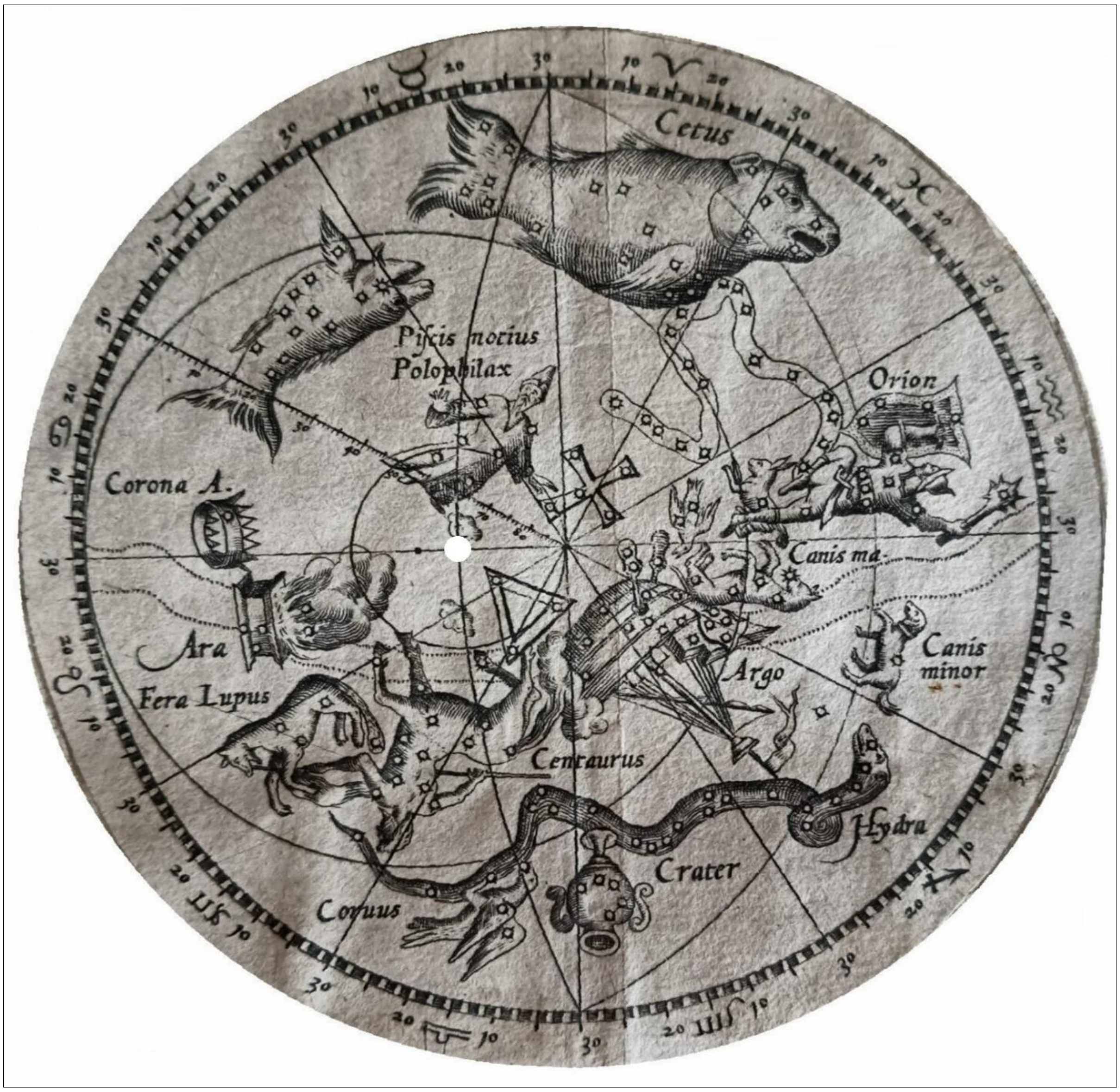

Figure 9: The Southern celestial hemisphere in Plancius' *Orbis Terrae Compendiosa Descriptio* (1596). The white circle shows the SCP (courtesy: private collection).

- The very purpose of explorations was to open new trade routes, and they could not be called scientific expeditions. The discoveries were primarily aimed at acquiring practical knowledge, useful for the success of the current and future voyages.
- The desire to provide an accurate cartographic representation of the starry sky had to wait a long time, and this seems to indicate that navigators, throughout the sixteenth century, could navigate the Southern seas with the knowledge they already possessed, taking care of the business that drove them to undertake long and dangerous voyages.
- In this context, Amerigo Vespucci was somewhat of an exception, because his intentions led him to see in his observations not only as a practical goal, but also a contribution, certainly defective, but not as defective as previously believed, to the growth of scientific knowledge. The origin of this sensitivity may perhaps be found in the fertile cultural climate of the city where he was born and educated.
- Previous attempts to interpret Vespucci's data were probably unsuccessful because they did not take into account the dual needs that he wanted to satisfy: to provide useful information to navigators and to expand the scientific knowledge available since antiquity, by a man possessing a high level of culture, but not specialized, at the beginning of the sixteenth century.

## 15 ACKNOWLEDGEMENTS

I thank the referees for their helpful comments.

**Dr. Davide Neri** was born in Cesena (Italy) in 1959. He received a Masters degree in Physics (1983) and a second one in Philosophy (1993) from the University of Bologna, and a PhD in Physics from the University of Bologna (1989) and a PhD in Historical and Philosophical Sciences from the University of Bergamo (2013).

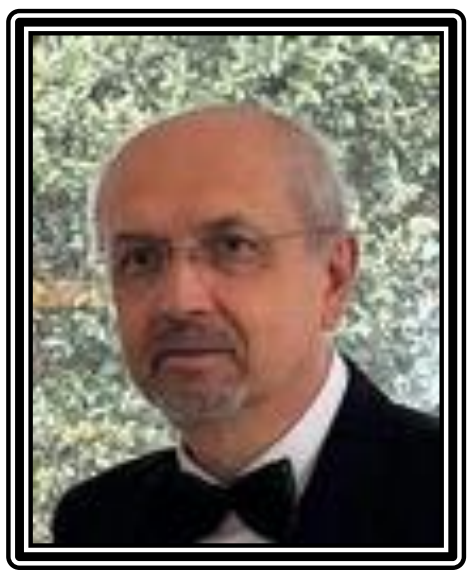

He taught Physics and Mathematics in secondary schools in the Province of Bologna for over 30 years, and was supervisor at the SSIS (school for teacher training) at the Universities of Bologna and Ferrara for 10 years. He concluded his teaching career at the Sabin High School in Bologna in 2021.

At present he is retired and continues to carry out research. Although he primarily worked as a teacher, he has published papers in the fields of the history of science, philosophy of physics, and history of celestial cartography, plus hypertexts and contributions to encyclopedias (e.g. the *Biographical Encyclopedia of Astronomers*).

His main interests have been the foundations of statistical thermodynamics, the history of electromagnetic theories in the nineteenth century, the transmission of knowledge from ancient times to the Renaissance, and the development of astronomical cartography between the fifteenth and eighteenth centuries. He has always been interested in scientific dissemination. His ORCID number is: 0009-0001-3160-7732.